\documentclass[aps,twocolumn,prl,superscriptaddress,longbibliography,reprint]{revtex4-2}

\usepackage{amsfonts}
\usepackage{mathrsfs}
\usepackage{amsmath}
\usepackage{color}
\usepackage{graphicx}
\usepackage{bm}
\usepackage{amssymb}
\usepackage{xspace}
\usepackage{epstopdf}
\usepackage{longtable}
\usepackage{multirow}
\usepackage{booktabs}
\usepackage{textcomp}
\usepackage[title]{appendix}
\usepackage{float}
\usepackage{dsfont}
\usepackage[colorlinks=true, letterpaper=true, pdfstartview=FitV, linkcolor=blue, citecolor=blue, urlcolor=blue]{hyperref}
\usepackage[]{changes}
\usepackage{overpic}

\providecommand{\U}[1]{\protect\rule{.1in}{.1in}}

\begin{document}

\title{Nonlinear Magnetic Orbital Hall Effect Induced by Spin-Orbit Coupling}

\author{Hui Wang}
\thanks{These authors contributed equally to this work.}
\affiliation{Science, Mathematics and Technology, Singapore University of Technology and Design, Singapore 487372, Singapore}

\author{Huiying Liu}
\thanks{These authors contributed equally to this work.}
\email{liuhuiying@buaa.edu.cn}
\affiliation{School of Physics, Beihang University, Beijing 100191, China}

\author{Yanfeng Ge}
\thanks{These authors contributed equally to this work.}
\affiliation{State Key Laboratory of Metastable Materials Science and Technology \& Hebei Key Laboratory of Microstructural Material Physics, School of Science, Yanshan University, Qinhuangdao, 066004, China}

\author{Xukun Feng}
\affiliation{Interdisciplinary Center for Theoretical Physics and Information Sciences (ICTPIS), Fudan University, Shanghai 200433, China}

\author{Jiaojiao Zhu}
\affiliation{LONGi Institute of Future Technology, and School of Materials \& Energy, Lanzhou University, 222 South Tianshui Road, Lanzhou 730000, China}

\author{Jin Cao}
\affiliation{Research Laboratory for Quantum Materials, Department of Applied Physics, The Hong Kong Polytechnic University, Kowloon, Hong Kong, China}

\author{Cong Xiao}
\email{congxiao@fudan.edu.cn}
\affiliation{Interdisciplinary Center for Theoretical Physics and Information Sciences (ICTPIS), Fudan University, Shanghai 200433, China}

\author{Shengyuan A. Yang}
\email{shengyuan.yang@polyu.edu.hk}
\affiliation{Research Laboratory for Quantum Materials, Department of Applied Physics, The Hong Kong Polytechnic University, Kowloon, Hong Kong, China}

\author{Lay Kee Ang}
\email{ricky\_ang@sutd.edu.sg}
\affiliation{Science, Mathematics and Technology, Singapore University of Technology and Design, Singapore 487372, Singapore}

\begin{abstract}
Electrical readout of 180$^\circ$ switching in strictly compensated collinear antiferromagnets remains a major challenge in antiferromagnetic spintronics. Electrical writing of perpendicularly magnetized ferromagnets by out-of-plane orbital torque remains an important challenge in orbitronics. In this work, we propose a second-order nonlinear magnetic orbital Hall effect in the source antiferromagnet as a simultaneous recipe for both difficulties. 
This orbitronics effect is induced by spin-orbit coupling and is odd in the N\'eel vector, thus is a unique effect that integrates both functionalities via electric control of the N\'eel vector in the source antiferromagnet.
Our first-principles calculations in CuMnAs predict significant non-perturbative orbital effects from spin-orbit coupling, with a orbital Berry-curvature dipole mechanism. These findings unveil new possibilities opened by topological antiferromagnetic orbitronics.
\end{abstract}
\maketitle

Orbitronics aims to utilize the orbital degree of freedom of electrons for carrying information. A focal ingredient is the orbital Hall effect, where an applied electric field drives a transverse flow of orbital angular momentum \cite{bernevig2005orbitronics,Guo2005,kontani2009giant,go2018intrinsic,Bhowal2020,sala2022giant,Salemi2022PRM,Manchon2022OHE,Rappoport2023orbital,Mertig2024OHE,Mertig2025OHE}. The orbital Hall effect in nonmagnetic materials can appear in the absence of spin-orbit coupling (SOC), which enables generating orbital current orders of magnitude surpassing the spin Hall current in weakly spin-orbit coupled materials \cite{jo2018gigantic}, attracting intensive recent interests \cite{choi2023observation,Kawakami2023OHE-Cr,DingPRL2024,Jiang2024OT-PMA-Zr,Klaui2025OT-PMA-Ru,Jiang2025orbital-FGT}. However, controlling and tuning the orbital Hall effect in nonmagnetic materials remain difficult. On the other hand, magnetic materials possess a characteristic magnetic orbital Hall effect (MOHE) \cite{Salemi2022MOHE}, which is odd in magnetization hence can be manipulated by the reorientation of the latter.
In particular, electrical manipulations of magnetic order parameters in ferromagnets and in compensated collinear antiferromagnets (AFM) have been experimentally demonstrated and theoretically well understood \cite{manchon2019}, rendering suitable platforms for exploring MOHE and the resultant electrically manipulable orbital torque exerted on adjacent target magnets.  

Compared to ferromagnets, AFMs enjoy the advantages of symmetry-enforced strictly zero net magnetization and high-speed magnetic dynamics that are favored by next-generation spintronic devices. However, MOHE in AFMs has not been studied yet.
In collinear AFMs, the two spin-sublattices are connected by combined time reversal ($\mathcal{T}$) and translation ($t$) or by combined $\mathcal{PT}$ symmetry ($\mathcal{P}$ is inversion).
Because the MOHE is forbidden by $\mathcal{T}t$ symmetry at any order of electric field and by $\mathcal{PT}$ at odd orders of field, a new orbital transport effect, namely the second-order nonlinear MOHE in $\mathcal{PT}$-symmetric AFMs, offers a unique opportunity for exploration. How does this new effect behave in generating out-of-plane orbital torque is an open question, which is a key issue for orbitronics information writing into perpendicular magnets \cite{Ryu2020SOT,miron2011,Liu2012,Yang2023briefing}, especially considering that out-of-plane orbital torque has not been realized in experiments by nonmagnetic orbital sources.

Noticeably, in collinear magnetic phases, the appearance of MOHE requires nonzero SOC \cite{feng2026MOT}, in sharp contrast to the case in nonmagnetic materials. On one hand, this dependence makes nonlinear MOHE a unique orbitronics effect that can detect the reorientation and 180$^\circ$ reversal of N\'eel vector in $\mathcal{PT}$ AFMs, which remains a central obstacle in AFM spin-orbitronics, alternative to the way of nonlinear charge transport \cite{Godinho2018Electrically,wang_intrinsic_2021,liu_intrinsic_2021}. On the other hand, the requirement on SOC poses the question as to whether or not the nonlinear MOHE can still dominate over its spin counterpart, i.e., nonlinear magnetic spin Hall effect (MSHE). If yes, whether or not strong SOC is necessary? The answers to these questions are critical for the prospect of nonlinear MOHE and AFM orbitronics.

In this work, we propose the nonlinear MOHE with the mechanism of orbital Berry-curvature dipole (OBD), and clarify its symmetry characters in $\mathcal{P}\mathcal{T}$-symmetric AFMs. We reveal its utility in electrical writing into the target perpendicular magnets by showing that it enables generating the desired unconventional out-of-plane collinearly polarized orbital current (CPOC) in 90$\%$ of all the possible magnetic point groups for $\mathcal{P}\mathcal{T}$ AFMs and identifying a wide range of material classes where the nonlinear MOHE is the leading origin of out-of-plane CPOC. By using the prototypical $\mathcal{P}\mathcal{T}$ AFM orthorhombic CuMnAs as an example, our first-principles calculations exhibit sizable CPOC from nonlinear MOHE, with a $2\pi$ periodicity on the N\'eel vector. 
These results not only indicate the desired $in$-$situ$ tuning of the out-of-plane orbital torque from nonlinear MOHE by electric control of AFM topological band structures via N\'eel torque, but also offer a novel orbitronics tool for electrically reading the N\'eel-vector 180$^{\circ}$ reversal in the source AFM itself.

Notably, even for orbital (spin) polarization of the generated orbital (spin) current parallel to the N\'eel vector, the nonlinear MOHE arising from SOC is found to be two orders of magnitude larger than the nonlinear MSHE that can appear in the zeroth order of SOC. We attribute this result to the small SOC around the Fermi level of CuMnAs, which amplifies the orbital effect by slightly gapping nodal lines. These features unveil that the nonlinear MOHE in topological AFMs is a non-perturbative effect of SOC, which requires SOC but can be significant even under weak SOC.


{\renewcommand{\arraystretch}{1.42}
\setlength\tabcolsep{6.6pt} 
\begin{table*}[hbt]
\centering
\caption{Symmetry constraints on CPOC from nonlinear MOHE and $\mathcal{T}$-even linear orbital Hall effect. $S_{3}^{z}$, $S_{6}^{z}$, $C_{3}^{z}\mathcal{T}$ and $C_{6}^{z}\mathcal{T}$ forbid all listed elements, hence are not shown.}
\begin{tabular}{ccccccccccccc}
\hline \hline
 & $\mathcal{P},S_{6}^{x}$ & $E$ & $M_{z}$ & $M_{x}$, $S_{3}^{x}$ & $M_{y}$ & $C_{2}^{z}$ & $C_{2}^{x}$ & $C_{2}^{y}$ & $C_{3,4,6}^{z}$ & $S_{4}^{z}$ & $C_{3}^{x}$ & $C_{4,6}^{x},S_{4}^{x}$\tabularnewline
 & $\mathcal{T},C_{3}^{x}\mathcal{T}$ & $\mathcal{PT}$ & $C_{2}^{z}\mathcal{T}$ & $C_{2}^{x}\mathcal{T},C_{6}^{x}\mathcal{T}$ & $C_{2}^{y}\mathcal{T}$ & $M_{z}\mathcal{T}$ & $M_{x}\mathcal{T}$ & $M_{y}\mathcal{T}$ & $S_{3,4,6}^{z}\mathcal{T}$ & $C_{4}^{z}\mathcal{T}$ & $S_{6}^{x}\mathcal{T}$ & $C_{4}^{x}\mathcal{T},S_{3,4}^{x}\mathcal{T}$\tabularnewline
\hline 
$\chi_{zxx}^{z}$ & $\times$ & $\surd$ & $\times$ & $\times$ & $\times$ & $\surd$ & $\surd$ & $\surd$ & $\surd$ & $\surd$ & $\surd$ & $\surd$\tabularnewline
$\chi_{z(xy)}^{z}$ & $\times$ & $\surd$ & $\times$ & $\surd$ & $\surd$ & $\surd$ & $\times$ & $\times$ & $\times$ & $\surd$ & $\surd$ & $\times$\tabularnewline
$\chi_{zyy}^{z}$ & $\times$ & $\surd$ & $\times$ & $\times$ & $\times$ & $\surd$ & $\surd$ & $\surd$ & $\chi_{zxx}^{z}$ & $-\chi_{zxx}^{z}$ & $\surd$ & $\surd$\tabularnewline
$(\sigma^{\text{int}})_{zx}^{z}$ & $\surd$ & $\surd$ & $\times$ & $\surd$ & $\times$ & $\times$ & $\surd$ & $\times$ & $\times$ & $\times$ & $\surd$ & $\surd$\tabularnewline
$(\sigma^{\text{int}})_{zy}^{z}$ & $\surd$ & $\surd$ & $\times$ & $\times$ & $\surd$ & $\times$ & $\times$ & $\surd$ & $\times$ & $\times$ & $\surd$ & $\times$\tabularnewline
\hline \hline 
\end{tabular}
\label{tab_sym}
\end{table*}}

\begin{figure*}[hbt]
\centering
\includegraphics[width=0.9\textwidth]{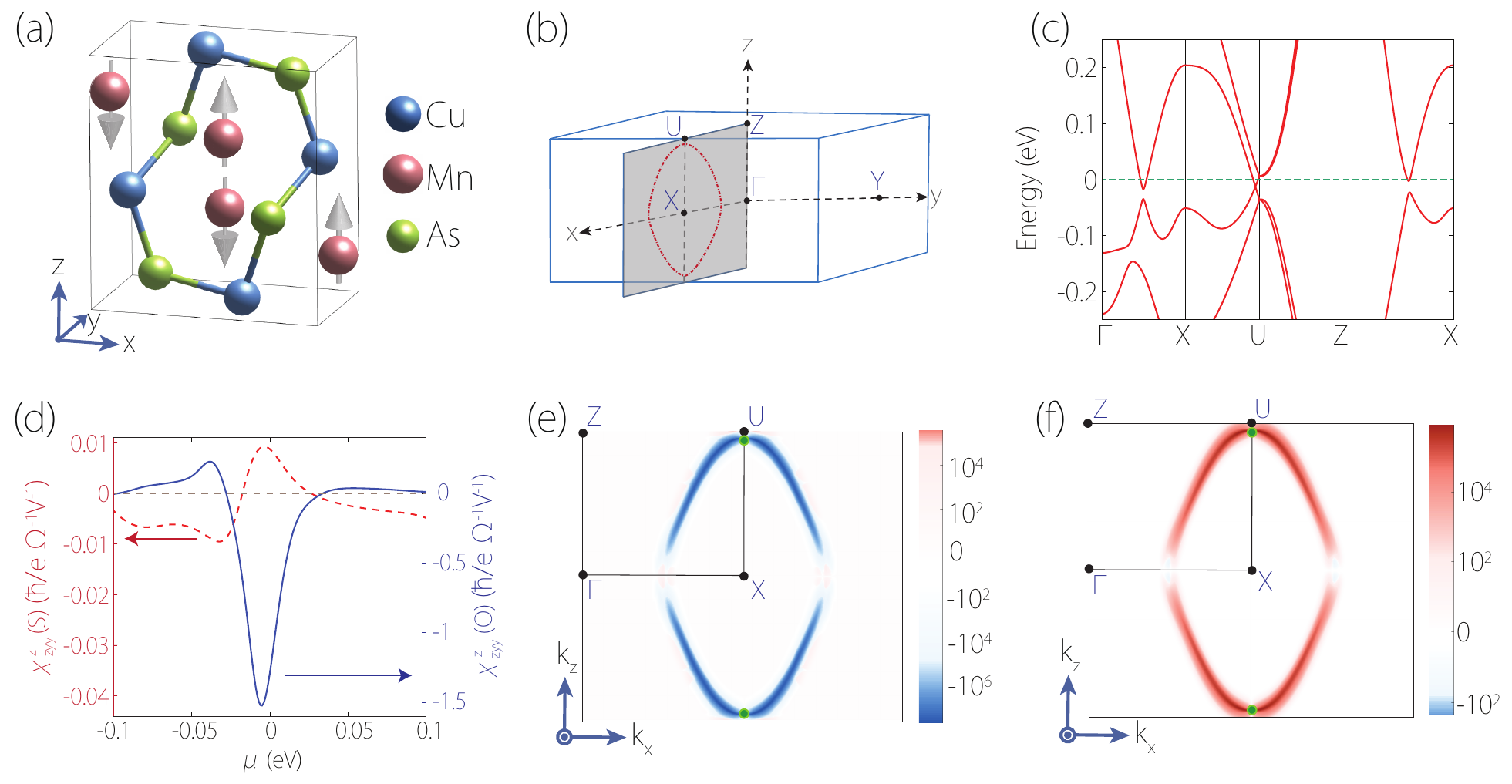}
\caption{(a) Lattice structure of AFM orthorhombic CuMnAs with N\'{e}el vector $\bm n\parallel[001]$. (b) Brillouin zone of CuMnAs. The red dotted curve illustrates the nodal line in the $k_y=0$ plane around X point obtained without SOC. (c) Calculated band structure of CuMnAs [001] including SOC. (d) CPOC from nonlinear magnetic orbital and spin Hall conductivity  $\chi_{zyy}^z$ as a function of chemical potential at $T=50$ K. (e,f) Distribution of $k$-resolved (e) $D_{zyy}^z(\text{O})$ and (f) $D_{zyy}^z(\text{S})$ on the intrinsic Fermi surface in the $k_y=0$ plane around the X point. The two green dots on the X-U line indicate the pair of Dirac points, which are related by $C_{2x}$.
}
\label{fig_CuMnAs}%
\end{figure*}

\textit{Symmetry properties and mechanism}--We first show that the nonlinear MOHE is a generic effect in $\mathcal{PT}$ AFMs from the symmetry perspective.
The second order orbital current response is expressed as
$\delta j_{a}^d=\chi_{abc}^dE_{b}E_{c}$,
where the current is along $a$ direction with $d$ polarization of orbital angular momentum. The roman indices label the Cartesian components, and the Einstein summation convention is adopted. As we are considering $\mathcal{T}$-odd transport, $\chi_{abc}^{d}$ is a $\mathcal{T}$-odd rank-4 pseudotensor ($\mathcal{T}$-odd or $\mathcal{T}$-even means the tensor flips its sign or not under $\mathcal{T}$ operation). The constraints imposed by the magnetic
point group symmetries follow the relation
\begin{equation}
\chi_{abc}^{d}=\eta\det(\mathcal{R})\mathcal{R}_{dd'}\mathcal{R}_{aa'}\mathcal{R}_{bb'}\mathcal{R}_{cc'}\chi_{a'b'c'}^{d'}\label{eq:symmetry}
\end{equation}
for $\mathcal{R}$ or $\mathcal{R}\mathcal{T}$ magnetic point group symmetry operations. Here $\mathcal{R}$ is the orthogonal transformation matrix for the crystalline symmetry. $\eta$ is taken as 1 for the nonprimed $\mathcal{R}$ operations, 
and $-1$ for the primed $\mathcal{R}\mathcal{T}$ combined operations.
From Eq.~(\ref{eq:symmetry}), $\chi_{abc}^{d}$ can exist in systems where $\mathcal{T}$ and $\mathcal{P}$ are both broken while the combined $\mathcal{P}\mathcal{T}$ preserves. We further find $\chi_{abc}^{d}$ is allowed in all 21 magnetic point groups with $\mathcal{P}\mathcal{T}$ symmetry but without $\mathcal{P}$ or $\mathcal{T}$ (see the group list in Table. S1 of Supplemental Material \cite{supp}), which constitute a large class of compensated collinear AFMs. In this class, the linear $\mathcal{T}$-odd MOHE and nonlinear $\mathcal{T}$-even OHE are forbidden by $\mathcal{P}\mathcal{T}$, and only the linear $\mathcal{T}$-even OHE and $\mathcal{T}$-odd nonlinear MOHE can be allowed. It is to be noted that since the global inversion symmetry of the two sectors is preserved in $\mathcal{P}\mathcal{T}$ AFMs, the crystal structure of the system is centrosymmetric, which is distinct from the intrinsic NOHE~\cite{wang2025NOHE} in nonmagnetic systems, where the global inversion symmetry must be broken.

We next analyze the symmetry characters of the CPOC induced by nonlinear MOHE. The pertinent response tensor elements read as $\chi_{abc}^{a}$ with $a\neq (b,c)$, and $b,c$ can be symmetrized by $\chi_{a(bc)}^{a}=(\chi_{abc}^{a}+\chi_{acb}^{a})/2$. Our analysis shows that such CPOC can be supported by 19 out of the 21 magnetic point groups above, except for 4/m$'$mm and 6/m$'$mm (details shown in Table S1). For m$'$mm, m$'$m$'$m$'$, $4'$/m$'$m$'$m, 4/m$'$m$'$m$'$, $6'$/mmm$'$, 6/m$'$m$'$m$'$, m$'\Bar{3}'$, m$'\Bar{3}'$m, and m$'\Bar{3}'$m$'$ particularly, the CPOC is forbidden in the linear effect, and $\chi_{abc}^{a}$ serves as the leading order effect. For the other magnetic point groups, CPOCs inspired by the linear and nonlinear effects may flow along different directions. In Table \ref{tab_sym}, we specify the driving-current plane as the $xy$ plane and list the constraints on the out-of-plane CPOC by all magnetic point-group operations. Table \ref{tab_sym} and Table S1 provide a complete magnetic point-group description for nonlinear MOHE, rendering a symmetry guidance for material research on this effect.

To have microscopic understanding of the nonlinear MOHE, we proceed to the physical mechanism of this effect. Notice that the orbital current does not change sign under $\mathcal{T}$ operation but the charge current does. As such, the physical mechanisms of $\mathcal{T}$-odd nonlinear MOHE are parallel to those of $\mathcal{T}$-even nonlinear charge Hall effect. For the latter, there are Berry curvature dipole \cite{fu2015,ma2019}, skew scattering \cite{kang2019,du2019}, side jump \cite{du2019} and other delicate disorder induced mechanisms \cite{Xiao2019NLHE,lu2021}. For material research, the Berry curvature dipole is the most focused mechanism, because only it is amenable to first-principles calculations and renders a benchmark for comparing theory and experiment. Therefore, in the nonlinear MOHE, we will focus on the orbital counterpart of Berry curvature dipole, which allows quantitative assessment for real materials.

Bloch electrons possess an anomalous velocity triggered by $E$ field and quantified by Berry curvature \cite{chang_berry_1995,xiao2010}. This anomalous velocity, when integrated over the nonequilibrium distribution driven by $E$, yields the Berry curvature dipole current in the $E^2$ order \cite{fu2015,ma2019}. Similarly, an anomalous orbital current of each Bloch electron labeled by ($n$, $\bm k$) reads $\delta j^{d,n}_a(\bm k)= \Omega_{ab}^{d,n}(\bm k)E_b$, where the underlying quantity is the orbital Berry curvature \cite{Bhowal2020,Manchon2022OHE,Rappoport2023orbital}
$
\Omega_{ab}^{d,n}(\bm k)=-2\operatorname{Im}\sum_{n^{\prime}\neq n}\big[j_{a}^{d,nn^{\prime}}(\bm k)v_{b}%
^{n^{\prime}n}(\bm k)/\omega^{2}_{nn'}(\bm k)\big].
$
Here, we set $e=\hbar=1$, $\omega_{nn'}=\varepsilon_{n}-\varepsilon_{n'}$ is the interband energy difference, and the numerator involves interband elements of orbital current and velocity operators.
Summing $\delta j^{d,n}_a(\bm k)$ over the nonequilibrium distribution $\delta f=\tau E_c\partial_cf_0$ ($f_0$ as the equilibrium Fermi-Dirac distribution, and $\partial_{c}\equiv\partial_{k_{c}}$) yields a quadratic orbital current density with
$ \chi_{abc}^{d}=\tau \mathcal{D}_{abc}^d$.
Here, $\tau$ is the relaxation time, and 
\begin{equation}
\mathcal{D}_{abc}^d=\sum_n\int\frac{d^\mathfrak{d} \bm k}{(2\pi)^\mathfrak{d}}\Omega_{ab}%
^{d,n}(\bm k)\partial_{c}f_{0}(\bm k)\label{both}%
\end{equation}
is the OBD, with $\mathfrak{d}$ as the dimensionality of the system.

The above rationale applies to the nonlinear MSHE in parallel. The difference between orbital and spin responses lies in their dependence on SOC.
The orbital and spin current operators ($\hat{j}_{a}^{d}$) can be expressed as \cite{sinova2015} $\frac{1}{2}\{ \hat{v}_{a},\hat{L}^{d}\}$ and $\frac{1}{2}\left\{ \hat{v}_{a},\hat{s}^{d}\right\}$, respectively, where the matrix elements of orbital angular momentum are given by \cite{Dai2020gfactor,Souza2023multipole,Mertig2024OHE,Mertig2025OHE}
\begin{equation}
\boldsymbol{\mathcal{L}}_{mn}=\frac{i}{4\mu
_{B}}\sum_{\ell\neq m,n}(\frac{1}{\varepsilon_{\ell}-\varepsilon_{m}}+\frac{1}{\varepsilon_{\ell}-\varepsilon_{n}})\boldsymbol{v}^{m\ell}\times\boldsymbol{v}^{\ell n}.\label{L}%
\end{equation}
In the absence of SOC, the spin-only group symmetry $C'_2$, which represents a twofold spin-space rotation about an axis perpendicular to the N\'eel vector combined with $\mathcal{T}$, imposes an effective  $\mathcal{T}$ on the orbital response. Under this symmetry, the integrand of OBD is an odd function of $\bm k$, thus the OBD vanishes. This null result highlights that OBD is a SOC-induced effect. In contrast, $C'_2$ does not suppress the spin counterpart of OBD \cite{Hayami2022}. On the other hand, the orbital moment is much enhanced and significant than spin around band near degeneracies. Therefore, even if weak SOC, when it induces a small local-gap around the Fermi surface, can trigger stronger nonlinear MOHE than MSHE. $\mathcal{PT}$ AFMs made up of weak-SOC elements thus provide possible platforms to explore this non-perturbative enhancement of nonlinear MOHE.

\textit{Application to CuMnAs}--
Combining our theory with first-principles calculations, we evaluate the OBD in the orthorhombic CuMnAs, which is a typical room temperature AFM metal with $\mathcal{PT}$ symmetry~\cite{MACA20121606,Emmanouilidou2017Magnetic,Zhang2017Massive,CaojinPhysRevLett2023}. The crystal structure is shown in Fig.~\ref{fig_CuMnAs}(a). Orthorhombic CuMnAs belongs to the space group $Pnma$ and point group $D_{2h}$. In the absence of SOC, the AFM state is described by the corresponding nonrelativistic spin space group symmetry (see Supplemental Material \cite{supp}). The low-energy bands exhibit a nodal line structure on the $k_y=0$ ($xz$) plane around the X point near the Fermi level [Fig.~\ref{fig_CuMnAs}(b)]. This nodal line structure is tied to the glide-mirror symmetry $R_y=\{m_y|(0,\tfrac{1}{2},0)\}$.
Once SOC is included, for CuMnAs with $\bm n\parallel[001]$, $R_y$ is no longer preserved, whereas the two-fold screw rotation $S_{z}=\{C_{2z}|(\frac{1}{2},0,\frac{1}{2})\}$ remains. As a result, the nodal line structure on the $k_y=0$ plane is gapped by SOC, and only the Dirac crossing on the X--U line survives due to the protection of $S_{z}$. The calculated band structure is shown in Fig.~\ref{fig_CuMnAs}(c), which is consistent with previous results~\cite{Tang2016Dirac,smejkal2017Electric,Huyen2021Spin}. 

CuMnAs [001] possesses the magnetic point group $m'm'm'$ that preserves $\mathcal{PT}$ and $S_{z}$ (see Table~S2). As our symmetry analysis has shown, this group forbids the linear $\mathcal{T}$-even OHE to provide CPOC, but allows CPOC from nonlinear MOHE. Take the $xy$ plane as the driving-field plane, there are two independent tensor elements of $\chi_{abc}^a$ for out-of-plane CPOC: $\chi_{zxx}^z$ and $\chi_{zyy}^z$. Figure~\ref{fig_CuMnAs}(d) presents the calculated $\chi_{zyy}^z$ as a function of chemical potential at $T=50$ K. The relaxation time is $\tau=1.4$ ps, estimated by calculating the Drude weight and referring to the experimental longitudinal resistivity~\cite{Zhang2017Massive}. The response becomes strongly enhanced as $\mu$ lies close to the intrinsic Fermi level, reaching $\chi_{zyy}^z=-1.3$ $(0.0087)$ $\hbar/e$ $\Omega^{-1}$V$^{-1}$ for nonlinear magnetic orbital (spin) Hall conductivity. The orbital signal exceeds the spin one by more than two orders of magnitude. Such a significant effect is comparable to the ever reported strongest nonlinear CPOC given by nonmagnetic materials \cite{wang2025NOHE}.

To clarify the origin of the pronounced nonlinear orbital response, we present the $k$-resolved distribution of the OBD $D_{zyy}^z$ in Fig.~\ref{fig_CuMnAs}(e), and plot its spin counterpart in Fig.~\ref{fig_CuMnAs}(f) for comparison. The dominant weight is sharply distributed around the gapped nodal line region, and the magnitude of OBD is typically two orders of magnitude greater than its spin companion.
We have inspected that the local-gap values on the nodal line are mostly below 20 meV. Such small SOC-induced band splitting around Fermi level is the origin of the large OBD, which is a non-perturbative effect of SOC. Similar non-perturbative effects also appear in SOC-induced giant anomalous Hall \cite{Onoda2008,nagaosa2010,ye2018massive,kim2018large,liu2018giant} and anomalous Nernst transport~\cite{sakai2018giant,guin2019zero,sakai2020iron} in ferromagnetic topological nodal-line metals.

\begin{figure}[t]
\centering
\includegraphics[width=1\columnwidth]{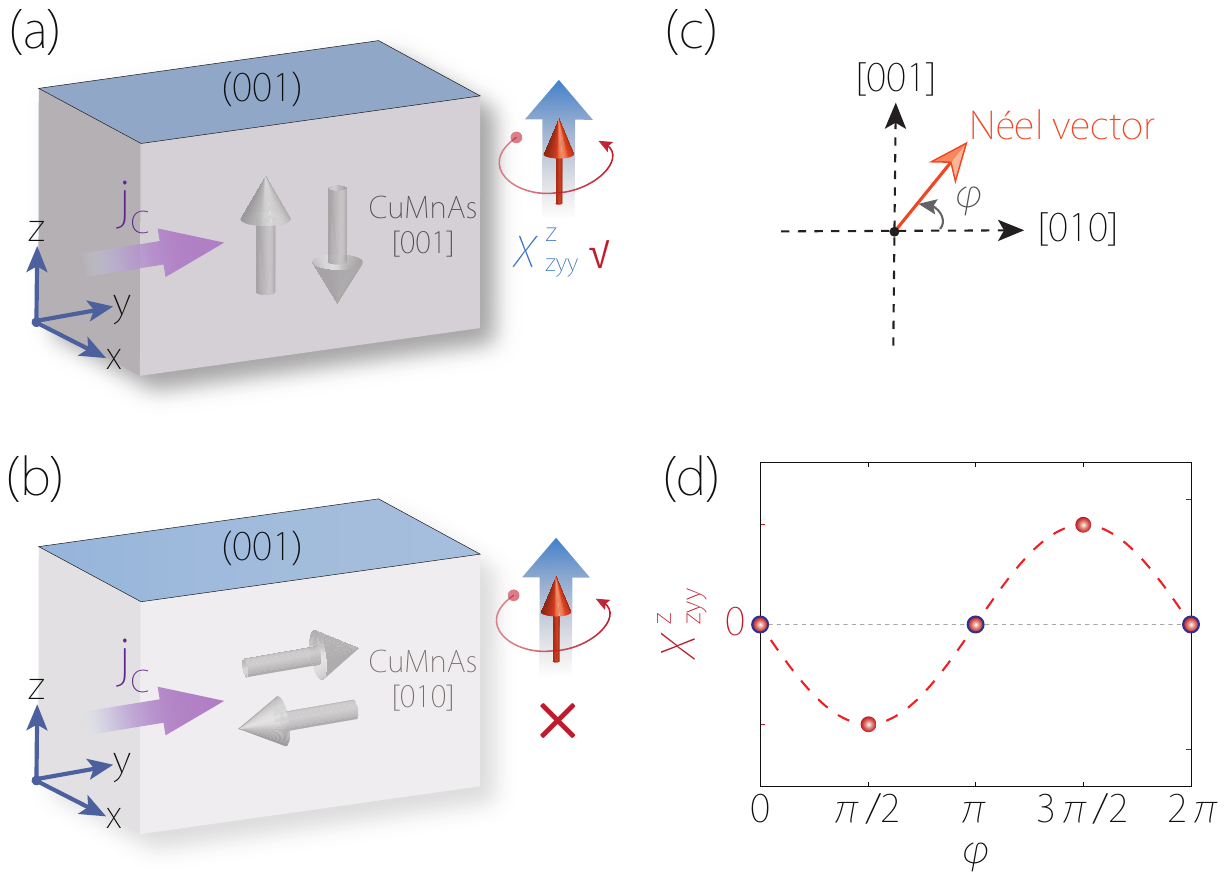}
\caption{Angular-dependent perpendicular CPOC from nonlinear MOHE in CuMnAs. (a) CuMnAs [001] and (b) CuMnAs [010] with $z$-axis as the out-of-plane direction. The $(001)$ plane is the contact interface to a target perpendicular ferromagnet. (c) Orientation of the N\'{e}el vector in the $yz$ plane, defined by the angle $\varphi$ measured from the [010] direction. (d) Angular dependence of the nonlinear orbital Hall conductivity $\chi_{zyy}^z$. The dashed line is a guide for eyes.}
\label{fig_device}%
\end{figure}

The interplay between magnetic ordering direction and CPOC provides a compelling avenue for controlling the nonlinear orbital torque generated from CuMnAs.  
When the (001) plane is the contact interface to the target ferromagnet, for $\bm n\parallel[001]$, a sizable out-of-plane CPOC via $\chi_{zyy}^z$ is allowed in response to a driving current in the $xy$ plane, as shown in Fig.~\ref{fig_device}(a), which generates an out-of-plane orbital torque on the adjacent ferromagnet. When the N\'{e}el vector is reoriented to the [010] direction \cite{Emmanouilidou2017Magnetic,Zhang2017Massive} by the current driving N\'eel spin-orbit torque on CuMnAs itself \cite{smejkal2017Electric,Shao2019}, the material possesses the magnetic point group $m'mm$ (Table S2), which does not permit out-of-plane CPOC (Fig.~\ref{fig_device}(b)). 
A shown in Fig.~\ref{fig_device}(d), a sizable $\chi_{zyy}^z$ appears for $\bm n\parallel[001]$, and its sign reverses upon flipping the N\'{e}el vector, whereas the response is suppressed for $\bm n\parallel[010]$ and $[0 \bar{1}0]$.
This variation in the nonlinear orbital Hall
conductivity can be detected by standard experimental methods
such as the magneto-optical Kerr effect \cite{Fan2014,Wang2015MOKE}.
Therefore, the nonlinear MOHE can be used to probe the reorientation and 180$^\circ$ reversal of N\'{e}el vector in CuMnAs.

\begin{figure*}[htb]
\centering
\includegraphics[width=1.7\columnwidth]{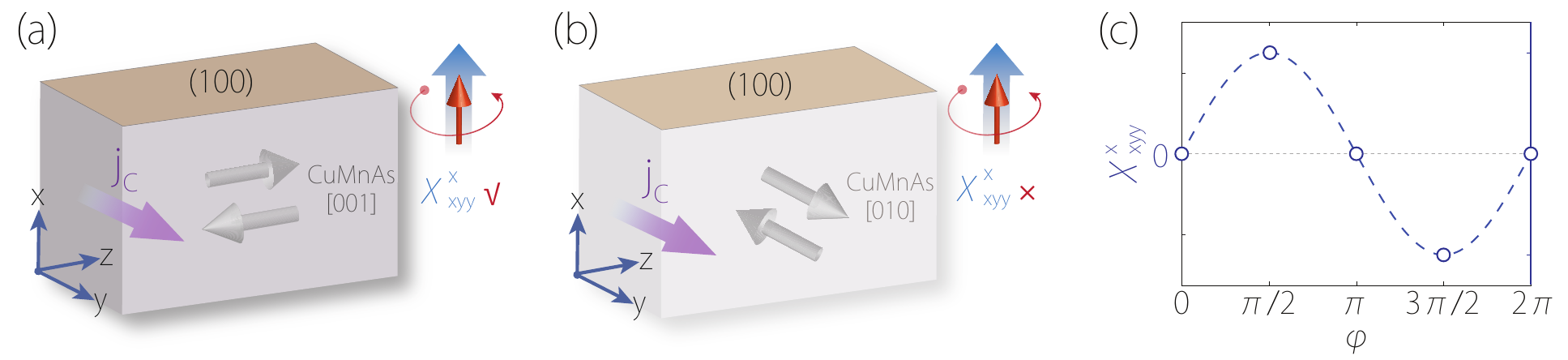}
\caption{Schematic of the orbit-torque device for (100) contact plane between CuMnAs and a perpendicular ferromagnet. (a) CuMnAs [001] and (b) CuMnAs [010] with $x$ axis as the out-of-plane direction. (c) Angular dependence of the nonlinear orbital Hall conductivity $\chi_{xyy}^x$. Here, $\varphi$ is defined in Fig.~\ref{fig_device}(c), and the dashed line is a guide for eyes.}
\label{fig_device_em}%
\end{figure*}


This functionality also takes effect for other choices of the interface between CuMnAs and the perpendicular ferromagnet, by using other CPOC components of the nonlinear MOHE. For example, when the $bc$ ($yz$) plane serves as the interface, the $x$ axis is the out-of-plane direction in the orbital-torque setup (Fig.~\ref{fig_device_em}(a) and (b)). For $\bm n\parallel[001]$, we find $\chi_{xyy}^x=1.4$ $(-0.0026)$ $\hbar/e$ $\Omega^{-1}$V$^{-1}$ for nonlinear magnetic orbital (spin) Hall conductivity at the Fermi level, generating a large out-of-plane CPOC by applying driving current in $y$ direction. When the N\'eel vector is reoriented to the [010] direction, the perpendicular CPOC is again prohibited, resulting in the angular dependence in Fig.~\ref{fig_device_em}(c).

\textit{Discussion}--
we have proposed the nonlinear MOHE in $\mathcal{PT}$-symmetric AFMs, which is odd in the N\'{e}el vector. Using first-principles calculations with OBD mechanism for orthorhombic CuMnAs, we find sizable nonlinear magnetic orbital Hall conductivities for the unconventional out-of-plane CPOC, and identify the origin in the SOC-gapped dispersive nodal line across the Fermi level. These results not only establish a new strategy for exploiting electrically controllable orbital current from strictly compensated collinear AFMs to achieve deterministic manipulation of the attached perpendicular ferromagnet, but also reveal an efficient orbitronics pathway for reading the AFM order itself. Our work uncovers unexplored potentials and utilities of topological AFM orbitronics emerging from intriguing interplay of nonlinear orbital transport and SOC.

From an application perspective, it is useful to estimate the effective orbital Hall conductivity in a representative heterostructure. The orbital Hall current injected into the ferromagnet is converted into a spin current~\cite{lee2021orbital,DingPRL2024}, with an
orbital-to-spin conversion efficiency $\eta$ depending on the ferromagnet. We can estimate an effective linear orbital Hall conductivity through $\sigma_{\mathrm{OH}}=\eta\chi \rho j$, where $j$ is the injected charge current density, and $\rho$ is the longitudinal resistivity. For $\chi_{zyy}^z$ of CuMnAs at 50 K, $j=\mathrm{10^7\ A/cm^2}$, and $\eta\sim 38\%$ in Fe$_3$GaTe$_2$~\cite{Jiang2025orbital-FGT}, we obtain $\sigma_{\mathrm{OH}} \approx 198\ \hbar/e\ \Omega^{-1}\mathrm{cm}^{-1}$. This value is much larger than those reported for linear out-of-plane spin current responses, such as $\mathrm{WTe_2}\sim 18$ $\hbar/e$ $\Omega^{-1}$cm$^{-1}$~\cite{Ralph2017WTe2} and $\mathrm{TaIrTe_4}\sim 103$ $\hbar/e$ $\Omega^{-1}$cm$^{-1}$~\cite{Yang2023TaIrTe4}, suggesting that CuMnAs/Fe$_3$GaTe$_2$ may provide a promising heterostructure platform for electrically controllable out-of-plane orbital torque on Fe$_3$GaTe$_2$ and for probing the N\'eel-vector 180$^\circ$ reversal in CuMnAs. The performance may be further enhanced by identifying materials analogous to CuMnAs with larger nonlinear responses and by optimizing the choice of heterostructures.

\bibliography{ref}

\bigskip
\onecolumngrid
\twocolumngrid
\vspace{0.5\baselineskip}

\appendix
\renewcommand{\theequation}{A\arabic{equation}} 
\setcounter{equation}{0} 
\renewcommand{\thefigure}{A\arabic{figure}}
\setcounter{figure}{0}



\end{document}